\pdfoutput=1
\documentclass[a4paper,twoside,
               showpacs,superscriptaddress,
               twocolumn,
               aps,prl,amsmath,amssymb]{revtex4}
\usepackage{graphicx}
\graphicspath{{figures/}}
\DeclareMathOperator{\re}{Re}

\begin{document}

\title{Increased Brownian force noise from molecular impacts in a constrained volume}

\author{A. Cavalleri}
  \affiliation{Centro Fisica degli Stati Aggregati, 38123 Povo,
    Trento, Italy}
\author{G. Ciani}
\author{R. Dolesi}
\author{A. Heptonstall}
  \affiliation{University of Glasgow, Glasgow G12 8QQ, UK}
\author{M. Hueller}
\author{D. Nicolodi}
  \email{nicolodi@science.unitn.it}
\author{S. Rowan}
  \affiliation{University of Glasgow, Glasgow G12 8QQ, UK}
\author{D. Tombolato}
\author{S. Vitale}
\author{P. J. Wass}
\author{W. J. Weber}
  \affiliation{Dipartimento di Fisica, Universit\`{a} di Trento, and
    INFN Gruppo di Trento, Via Sommarive 14, 38123 Povo, Trento, Italy}
 
\date{\today}

\begin{abstract}
  We report on residual gas damping of the motion of a macroscopic
  test mass enclosed in a nearby housing in the molecular flow regime.
  The damping coefficient, and thus the associated thermal force
  noise, is found to increase significantly when the distance between
  test mass and surrounding walls is smaller than the test mass
  itself. The effect has been investigated with two torsion pendulums
  of different geometry and has been modelled in a numerical
  simulation whose predictions are in good agreement with the
  measurements. Relevant to a wide variety of small-force experiments,
  the residual-gas force noise power for the test masses in the LISA
  gravitational wave observatory is roughly a factor $15$ larger than
  in an infinite gas volume, though still compatible with the target
  acceleration noise of
  $3\,\mbox{fm}\,\mbox{s}^{-2}\,\mbox{Hz}^{-1/2}$ at the foreseen
  pressure below $10^{-6}\,\mbox{Pa}$.
\end{abstract}

\maketitle

The fluctuation-dissipation theorem \cite{callen-1951} states that any
system with dissipation exhibits fluctuations analogous to Brownian
motion that can be modelled with an external driving force with power
spectral density
\begin{equation}
  S_F(\omega) = 4\, k\, T \re\left(\frac{\partial F}{\partial v}\right) = 4\, k\, T \re\bigl(Z(\omega)\bigr)
  \label{eq:fluctuation-dissipation}
\end{equation}
where $k$ is Boltzmann's constant, $T$ is the temperature of the
system, and $Z(\omega)$ is the mechanical impedance of the system.
Those equilibrium fluctuations are a fundamental limit of precision
metrology experiments employing macroscopic test-masses nominally in
perfect free-fall to define geodesic reference frames.

An unavoidable source of dissipation comes from residual gas in the
experimental volume, which yields a viscous, frequency independent,
impedance $Z(\omega) = \beta$. Theoretical calculations for small
force experiments~\cite{hinkle-1991, saulson-1990, schumaker-2003} and
direct observations~\cite{hagedorn-2006, numata-2007} show the gas
damping coefficient $\beta$ to be proportional to residual gas
pressure $P$ and to some effective surface area, which depends on the
test mass geometry.

A recent calculation~\cite{weber-2009} for a cubic test mass has found
that the translational damping coefficient is
\begin{equation}
  \beta_{\mathrm{tr}}^{\infty} = P\, s^2 \left(1+\frac{\pi}{8}\right) \left(\frac{32\, m}{\pi\, k\, T}\right)^{1/2}
  \label{eq:beta-tr}
\end{equation}
where $s$ is the side length of the test mass, and $m$ is the mass of
the gas molecules. For rotation, relevant for the experiments we will
present shortly, the coefficient is
\begin{equation}
  \beta_{\mathrm{rot}}^{\infty} = P\, s^4 \left(1 + \frac{\pi}{12}\right) \left(\frac{2\, m}{\pi\, k\, T}\right)^{1/2}
  \label{eq:beta-rot}
\end{equation}
This calculation assumes that gas and test mass are in thermodynamic
equilibrium, that gas molecule collisions with the test mass are
completely inelastic, with prompt stochastic re-emission from the
surface with a Maxwell-Boltzmann velocity distribution and cosine law
angular distribution. Importantly, this and other calculations assume
that the test mass is surrounded by an infinite volume of
collisionless gas, hence the superscript $\infty$, such that a
molecule emitted from the test mass surface disappears into the
surrounding volume, and that the momentum it imparts on the test mass
is uncorrelated with any subsequent collisions. This assumption breaks
down when the distance between the test mass and the surrounding
enclosure is only a fraction of the test mass size, as in the geometry
studied here, where the gap is roughly one tenth of the test
mass size.

Our experimental setup probes small forces relevant to the quality of
free fall achievable for the test masses of the Laser Interferometer
Space Antenna (LISA) gravitational-wave
observatory~\cite{bender-2000, vitale-2002} and its precursor LISA
Pathfinder~\cite{anza-2005}, using a torsion pendulum. External
force disturbances are suppressed to a level where the thermal noise
plays an important role in limiting the force sensitivity.
Measurements of the torsion pendulum free-motion damping show a
residual gas contribution several times larger than predicted by the
infinite-volume model, accompanied by an observed noise increase
consistent with mechanical thermal noise. We present here simple
arguments, based on the impedance to stochastic molecular flow through
the small gaps around the test mass, supported by numerical
simulation, that quantitatively explain this observed increased
dissipation. Enhanced gas damping has also been observed in
micro-mechanical resonators~\cite{zook-1992}, though the
interpretations have employed an elastic model of the molecule-wall
interaction~\cite{bao-2002}.

\begin{figure}[tb]
  \centering
  \includegraphics[width=0.9\columnwidth]{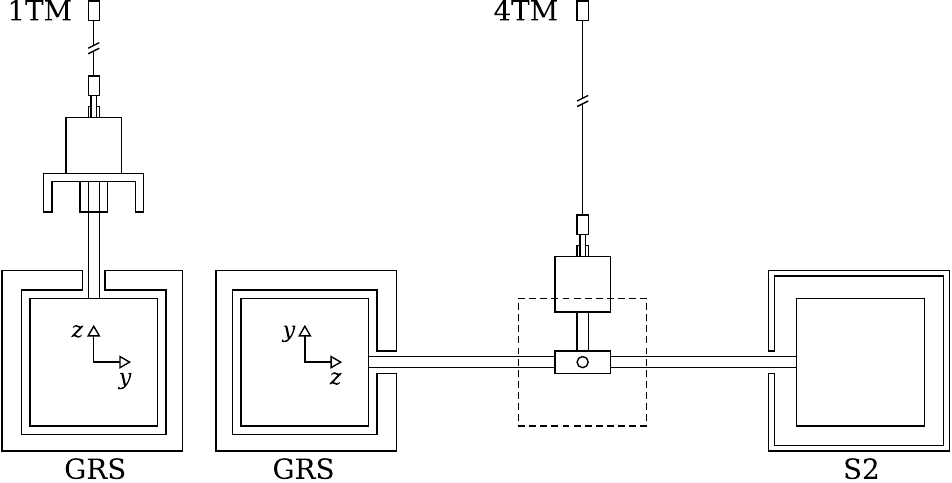}
  \caption{Sketch of the 1TM and 4TM pendulum inertial member
    surrounded by the GRS. Test mass side is $46\,\mbox{mm}$. Gaps
    between test mass and GRS are $4.0$, $2.9$, and $3.5$~mm on
    respectively the $x$, $y,$ and $z$ faces. The S2 gaps are $8.0$,
    $6.0$, and $8.0\,\mbox{mm}$ gaps on $x$, $y,$ and $z$ faces,
    respectively.}
  \label{fig:pendulums}
\end{figure}

Two different torsion pendulum geometries are used for the
experimental investigation. Both suspend a hollow replica of the LISA
Pathfinder cubic test mass with side length $s = 46\,\mbox{mm}$ inside
a prototype of the LISA Pathfinder Gravity Reference Sensor
(GRS)~\cite{dolesi-2003} that completely surrounds the test mass. All
GRS, test mass, and other inertial member surfaces are gold coated. In
the 1TM torsion pendulum~\cite{carbone-2007, cavalleri-2009a} the
rotation axis coincides with the test mass $z$ axis of symmetry, as
shown on the left of Fig.~\ref{fig:pendulums}. In the 4TM torsion
pendulum~\cite{carbone-2006, cavalleri-2009b} the test mass enclosed
by the GRS is one of four identical test masses in a cross shaped
configuration, suspended off-center with respect to the fiber axis
with a $r=0.1065\,\mbox{m}$ arm, as shown on the right of
Fig.~\ref{fig:pendulums}. The opposite test mass is enclosed in a
similar capacitive sensor (S2), featuring larger gaps. The 4TM
pendulum employs a $50\,\mu\mbox{m}$ diameter tungsten fiber, while
the 1TM pendulum uses a $40\,\mu\mbox{m}$ fused silica fiber with much
lower internal dissipation. Both torsion pendulums facilities operate
at the temperature of $293\,\mbox{K}$, controlled within
$0.1\,\mbox{K}$.

For both pendulums we have measured the torsional damping coefficient
in the pressure range of roughly $0.003$--$3\,\mbox{mPa}$ (mean free
paths of order $10\,\mbox{km}$ down to $10\,\mbox{m}$), by direct
measurement of the amplitude decay time $\tau$ of the free pendulum
motion \begin{equation}
  \beta = - \frac{\partial N}{\partial \dot{\varphi}} = \frac{2 \mathcal{I}}{\tau}
\end{equation}
where $\mathcal{I}$ is the torsion pendulum moment of inertia. The
decay time is computed as the average of the ratio between oscillation
amplitude, with each amplitude point averaged over three periods, and
its time derivative, as estimated by the difference between successive
points. Uncertainty is computed on statistical basis.

Measurements were typically performed over $1$--$3$ days, using large
initial amplitudes of order mrad. No amplitude dependence of $\tau$
was observed. The 1TM and 4TM oscillation periods were roughly $460$
and $1400\,\mbox{s}$, respectively, and their ring-down times in the
limit of zero pressure were roughly $6$ years ($Q \simeq 1.2\times
10^6$) and $16$ days ($Q \simeq 3.2 \times 10^3$). In the case of the
4TM torsion pendulum we performed the measurement both with and
without the GRS around the test mass.

The pressure in the torsion pendulum vacuum chambers is measured with
ion gauges cross-calibrated with a certified pressure gauge whose
accuracy is within 15\%, employing the calibration factor recommended
for air. While we have not as yet identified the dominant gas species
in our chamber, measurements with the same apparatus have confirmed
the pressure-dependent radiometric effect model~\cite{carbone-2007b,
  cavalleri-2009a} at the 10\% level, indicating similar systematic
uncertainty in the pressure measurements here.

\begin{figure}[tb]
  \centering
  \includegraphics[width=0.95\columnwidth]{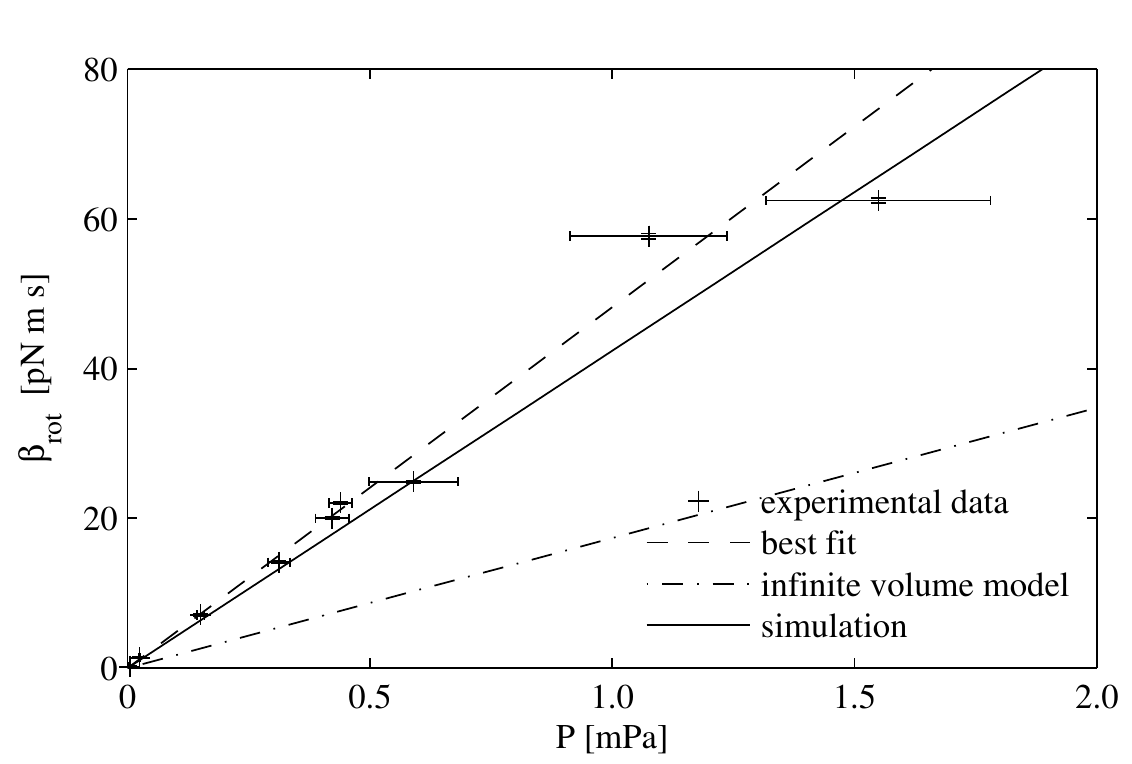}
  \caption{Measured 1TM gas damping coefficients
    $\beta_{\mathrm{rot}}$ as function of residual gas pressure $P$.
    Infinite-volume model and simulation predictions are also shown.}
  \label{fig:1tm-beta-pressure}
\end{figure}

\begin{figure}[tb]
  \centering
  \includegraphics[width=0.95\columnwidth]{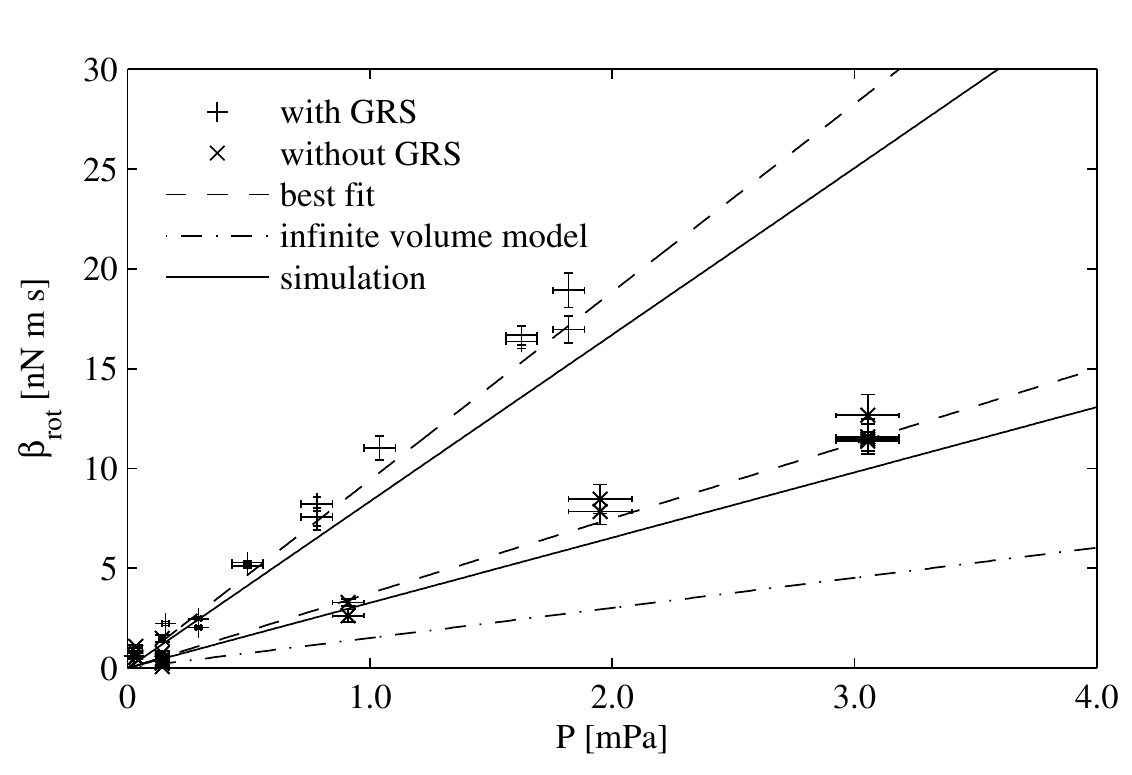}
  \caption{Measured 4TM gas damping coefficients
    $\beta_{\mathrm{rot}}$ as function of residual gas pressure $P$, with
    and without the GRS surrounding one of the test masses.
    Infinite-volume model and simulation predictions are also shown.}
  \label{fig:4tm-beta-pressure}
\end{figure}

The measured damping coefficients, plotted in
Figs.~\ref{fig:1tm-beta-pressure} and~\ref{fig:4tm-beta-pressure}, are
found to depend linearly on pressure. The residual zero-pressure
intercept, from other dissipation mechanisms, has been removed from
both datasets, and the pressure dependence is summarized in
Table~\ref{table:results}. The 1TM data can be directly compared with
the infinite volume prediction in Eqn.~\ref{eq:beta-rot}. The 4TM
prediction is calculated from the rotational and translational
contributions, $\beta = 4 \left(r^2\beta_{\mathrm{tr}}^{\infty} +
  \beta_{\mathrm{rot}}^{\infty}\right)$. The infinite-volume model
predictions are several times smaller than the measured values and
would foresee no difference caused by the removal of the GRS, while
here it lowers the total 4TM damping by more than a factor two.

\begin{table}[htb]
  \begin{tabular}{cccc}
    \hline
    \hline
    & measurement & $\infty$ model & simulation \\
    & $\partial\beta/\partial P\left[\mbox{m}^3\,\mbox{s}\right]$
    & $\partial\beta/\partial P\left[\mbox{m}^3\,\mbox{s}\right]$
    & $\partial\beta/\partial P\left[\mbox{m}^3\,\mbox{s}\right]$ \\    
    \hline
    I.   & $(4.8 \pm 0.1)\times 10^{-8}$
         & $1.74 \times 10^{-8}$
         & $4.24 \times 10^{-8}$ \\
    II.  & $(9.4 \pm 0.2)\times 10^{-6}$
         & $1.51 \times 10^{-6}$
         & $7.87 \times 10^{-6}$ \\
    III. & $(3.7 \pm 0.1)\times 10^{-6}$
         & $1.51 \times 10^{-6}$
         & $3.27 \times 10^{-6}$ \\
    IV.  & $(5.7 \pm 0.3)\times 10^{-6}$
         & $0$ 
         & $4.60 \times 10^{-6}$ \\
    \hline
    \hline
  \end{tabular}
  \caption{Comparison of $\beta$ pressure dependence, obtained from
    measurement, infinite-volume model, and simulation, in different
    configurations: I. 1TM, II. 4TM with GRS, III. 4TM without GRS, IV. 4TM GRS 
    contribution, obtained as difference between configurations II and
    III. Simulation uncertainty are on the fourth significative digit.}
  \label{table:results}
\end{table}

\begin{figure}[tb]
  \centering
  \includegraphics[width=0.95\columnwidth]{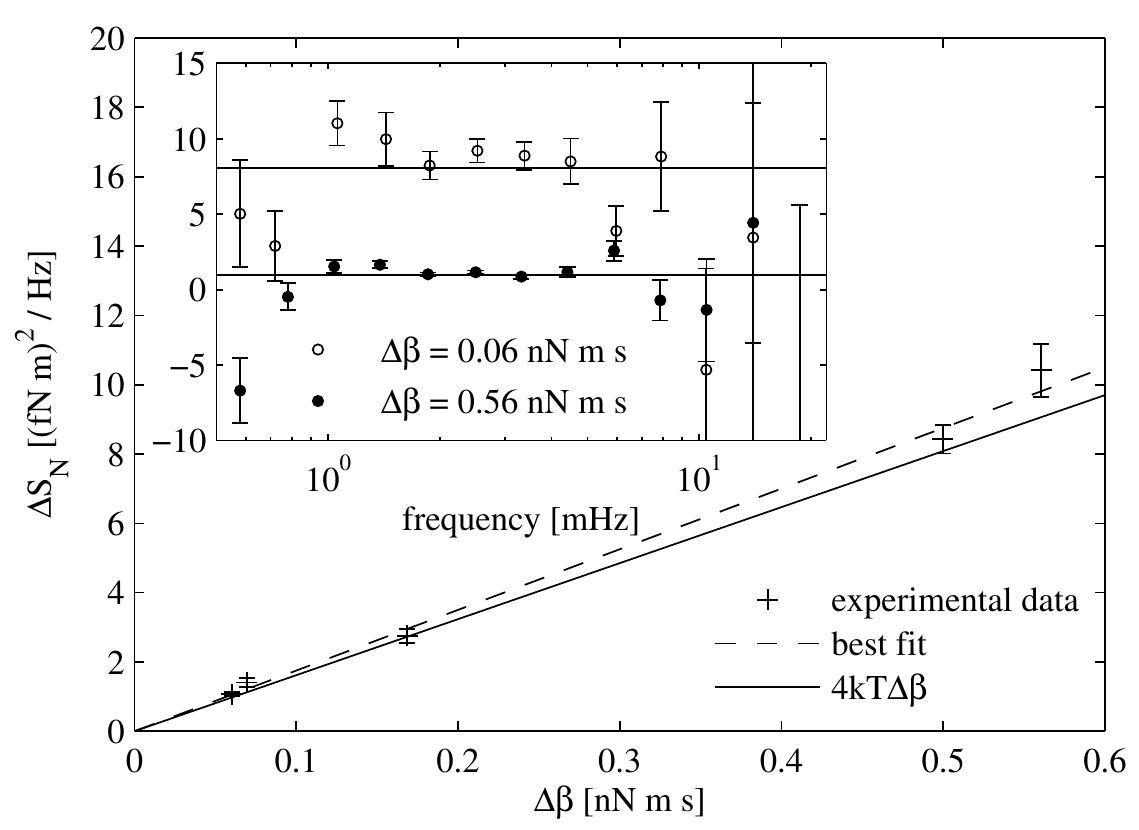}
  \caption{Measured averaged gas damping noise difference $\Delta S_N$
    in the band $0.5$--$12\,\mbox{mHz}$ as a function of differential
    damping coefficient $\Delta \beta$, compared with the Brownian
    noise prediction. Inset shows two noise spectra, compared against
    the predicted noise level.}
  \label{fig:noise-beta}
\end{figure}

A series of torque noise measurements, performed with the 1TM
pendulum, has confirmed the Brownian nature of the gas damping noise.
Measurements were performed at six different pressures, from $P_0 =
0.004$ to $15\,\mbox{mPa}$. Torque noise was estimated with a
dual-readout cross-correlation technique~\cite{carbone-2007,
  cavalleri-2009a}. For each measurement, lasting $1$--$3$ days, noise
spectra were averaged over 8 to 20 windows of $25000\,\mbox{s}$ each,
with 50\% overlap and over logarithmically-spaced frequency bins, with
uncertainties estimated as the standard deviation divided by the
square root of the number of points. To identify the gas damping
noise, the $P_0$ spectrum was subtracted from the five higher pressure
measurements, yielding a differential spectrum $\Delta S_N$ for each
$\Delta\beta \equiv \beta(P) - \beta(P_0)$. Two such spectra are shown in
the inset of Fig.~\ref{fig:noise-beta}. The gas damping noise is best
fit by a white noise model between $0.5$ and $12\,\mbox{mHz}$, where
the gas damping noise is resolved. The averages of each differential
spectrum in this bandwidth are plotted as a function of $\Delta\beta$
in Fig.~\ref{fig:noise-beta}, with a linear fit yielding $\Delta S_N =
\left(1.08 \pm 0.03\right) \times 4\,k\,T\Delta\beta$. The gas damping
noise is thus white and within 10\% of the Brownian noise prediction.

The excess gas damping and associated force noise, well above that
foreseen by the infinite-volume model, is caused by the nearby GRS
walls, as can be understood by a simple picture of molecular gas flow.
Test mass motion along the $x$ direction, with velocity $v$, requires
gas flow $Q$, expressed in $\mbox{Pa}\,\mbox{m}^3\,\mbox{s}^{-1}$, from
the volume on one $x$ face to the other, to restore pressure
equilibrium. The finite molecular flow conductance $C$ of the channels
provided by the electrode-housing $y$ and $z$ faces gaps requires a
pressure drop $\Delta P$ to maintain this gas flow. In steady state
conditions, $Q = P s^2 v$ and thus the pressure drop must be $\Delta P
= Q / C = P s^2 v / C$. The test mass is thus subject to a force, $F =
\Delta P s^2$, proportional to the velocity $v$, and thus to an
increase in the viscous damping
\begin{equation}
  \beta = \left| \frac{\partial F}{\partial v} \right| \approx \frac{P s^4}{C}
  \label{eq:beta-model}
\end{equation}
Gas flow in a finite conductance is a dissipative process.

In an equivalent picture from the standpoint of the associated force
noise, the flow impedance of the channels is related to the time
needed for molecules to random walk from one side of the test mass to
the other. This random walk requires repeated impacts of a single
molecule on the same test mass face and thus introduces correlations
between subsequent collisions, with repeated force impulses of the
same sign slowing the averaging out of the net force and thus
increasing force noise.

To verify our hypotheses, we developed, on the same assumptions as the
infinite-volume model, a numerical simulation of the gas dynamics
within a simplified geometry of a cubic test mass inside a cubic
housing. The gas is supposed to be composed of a single specimen with
molecular mass of $30\,\mbox{amu}$. The simulation is composed of $N$
steps, simulating each one the dynamics of a single molecule. The
simulation allows the system to evolve for a time $\Delta t$ chosen
long enough to allow time for the molecules to stochastically move
many times from one side of the test mass to the other. Momentum
exchange between test mass and gas molecules is recorded at each
collision. The gas-damping coefficient $\beta$ is computed from the
variance $\sigma^2_F$ of each component of the total force acting on
the test mass: \begin{equation}
  \beta_{\mathrm{tr}}^{\mathrm{sim}} = \frac{S_F}{4\,k\,T} = \frac{\Delta t}{2\, k\, T}\,\sigma^2_F
\end{equation}
An equivalent approach is used to calculate
$\beta_{\mathrm{rot}}^{\mathrm{sim}}$ from the mean square fluctuation
of the torque.

\begin{figure}[tb]
  \centering
  \includegraphics[width=0.95\columnwidth]{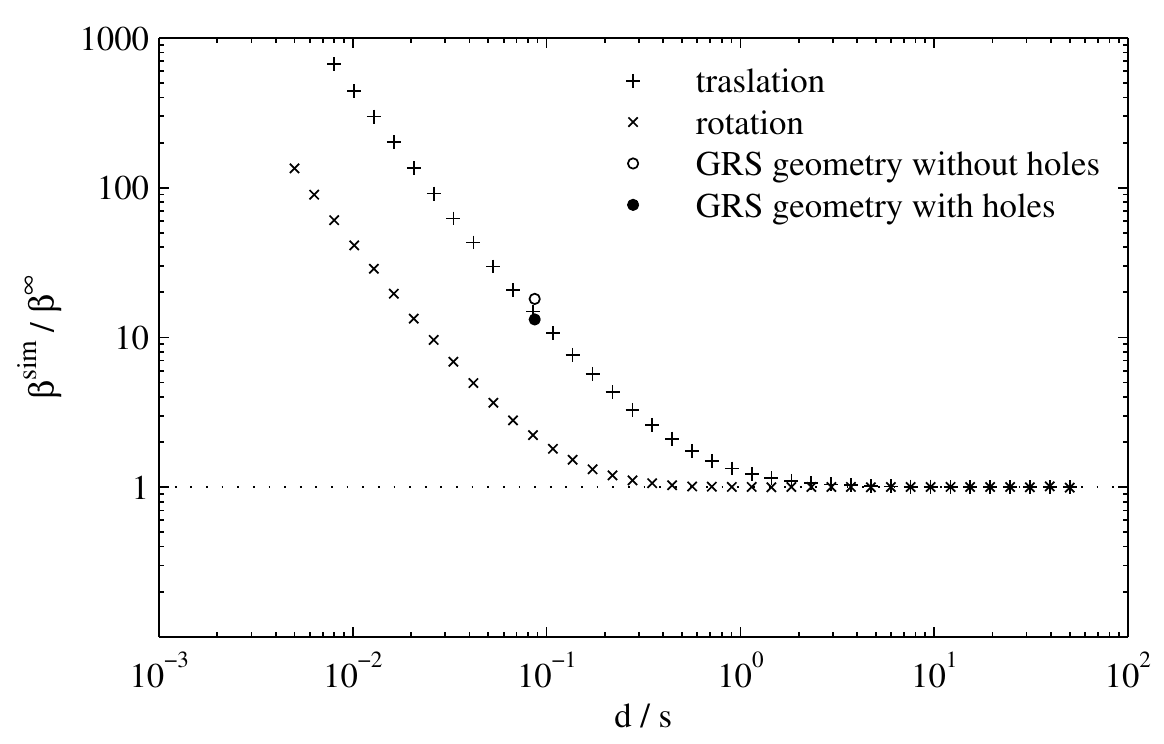}
  \caption{Gas damping $\beta^{\mathrm{sim}}$ obtained from the
    numerical simulation for different test mass side lengths $s$ and
    gap sizes $d$, normalized to the infinite-volume model prediction
    $\beta^{\infty}$.}
  \label{fig:simulation}
\end{figure}

The simulation has been performed as a function of gap size, with the
results shown in Fig.~\ref{fig:simulation}. In addition to confirming
the analytic model in Eqns.~\ref{eq:beta-tr} and~\ref{eq:beta-rot} for
large gaps, the data show the expected increased damping for $d
\lesssim s$, which approaches a power law $(d/s)^{-2}$ for vanishing
gap. This can be roughly understood with the flow impedance arguments
presented earlier. For $s \gg d$, the channel conductance is of order
$C \approx d^2 \ln\left(s/d\right)\left(k
  T/m\right)^{1/2}$~\cite{livesey-2007}. Substituting into
Eqn.~\ref{eq:beta-model} and comparing with Eqn.~\ref{eq:beta-tr}
\begin{equation}
  \beta_{\mathrm{tr}} \approx
  \frac{\beta_{\mathrm{tr}}^{\infty}}{ \ln\left(s/d\right)\left(d/s\right)^{2}}
\label{func_depend}
\end{equation}
This asymptotically approaches a log slope of $-2$ for vanishing gap,
as observed in the simulation data.

Fig.~\ref{fig:simulation} also shows simulation results obtained
simulating the slightly asymmetric gaps of the real GRS geometry.
Finally, we also simulate the $6\,\mbox{mm}$ diameter holes present on
the $x$ faces, which act as vents that reduce the damping pressure
head, with a roughly 20\% reduction of the gas damping obtained in the
simulation.

The measured 1TM gas damping coefficient can be directly compared with
the simulation results, adding in small contributions due to minor
features of the inertial member computed with the infinite-volume
model. The gas damping coefficient for the 4TM torsion pendulum
without the GRS is computed from simulation results for the S2
geometry and from the infinite-volume model for the other three test
masses $\beta = 3\left(r^2\beta_{\mathrm{tr}}^{\infty} +
  \beta_{\mathrm{rot}}^{\mathrm{\infty}}\right) +
r^2\beta_{\mathrm{tr}}^{\mathrm{sim}} +
\beta_{\mathrm{rot}}^{\mathrm{sim}}$. The GRS contribution in the 4TM
gas damping coefficient is computed from simulation results and
infinite-volume model predictions $\beta =
r^2\left(\beta_{\mathrm{tr}}^{\mathrm{sim}} -
  \beta_{\mathrm{tr}}^{\infty}\right) +
\left(\beta_{\mathrm{rot}}^{\mathrm{sim}} -
  \beta_{\mathrm{rot}}^{\infty}\right)$. Results are summarized in
Table~\ref{table:results}. The agreement with measurements is within
20\%. This is good in respect of the approximations in the geometrical
model of the GRS contained in the simulation. We note that in the 4TM
pendulum the gas damping coefficient is dominated by the translational
part, the rotational one accounting for roughly 1\% of the total, and
thus the 4TM measurement validates the translational damping relevant
to force noise.

Converting the simulated residual-gas Brownian force-noise into
acceleration noise for LISA and LISA Pathfinder, with a $2\,\mbox{kg}$
test mass, yields
\begin{equation}
  S_a^{1/2} =  1.3 \times 10^{-15} \left(\frac{P}{10^{-6}\,\text{Pa}}\right)^{1/2}\;\mbox{m}\,\mbox{s}^{-2}\,\mbox{Hz}^{-1/2}\mbox{.}
  \label{eq:noise}
\end{equation}
Eqn.~\ref{eq:noise} is compatible with the target acceleration noise
of LISA Pathfinder and LISA, the latter of which will operate at
pressure below $10^{-6}\,\mbox{Pa}$, reached by venting the GRS to
space~\cite{bender-2000}. Gas damping is likely to be the dominant
pressure-related force noise for LISA. We note that, unlike many force
noise sources, Brownian noise from residual gas does not improve at
higher frequencies, this is important for future gravitational wave
missions, such as DECIGO and BBO, whose ambitious sensitivity goals
near $0.1\,\mbox{Hz}$ and shorter interferometry arms require even
lower acceleration noise limits \cite{decigo,bbo}. Finally, the
gas-damping question effectively confirms the need, also based on
electrostatic concerns \cite{dolesi-2003}, for gaps of at least
several mm: for a similar sized test mass and gaps of
$0.3\,\mbox{mm}$, roughly the gap employed in several space
accelerometry missions \cite{microscope, goce}, the resulting
acceleration noise from this source alone would be an order of
magnitude above the LISA goal.

\end{document}